\documentstyle[11pt]{article}

\title{\vspace{4cm}\large\bf
Discovering internal symmetry in cosmology}
\author{Arthur D.~Chernin\\
 Sternberg Astronomical Institute, Moscow University, Moscow, 119899,
Russia, \\
Astronomy Division, University of Oulu, FIN-90014,  Finland
}

\date{~}

\begin{document}

\maketitle

\begin{abstract}
\noindent
Internal (non-geometric) symmetry is recognized and studied
as a new phenomenon in cosmology.
Symmetry relates cosmic vacuum to non-vacuum forms of cosmic
energies, which are dark matter, baryons, and radiation.
It is argued that the origin and physical nature of cosmic internal symmetry
are due to the interplay between gravity and electroweak scale physics
in the early universe. The time-independent characteristic number associated
with this symmetry is found in terms of the Planck energy
 and the electroweak energy scale.
Cosmic internal symmetry provides a framework which suggests a
 consistent solution to both the
`cosmic coincidence problem' and the `naturalness problem' that
have been considered as the most severe challenge to current concepts in
cosmology and fundamental physics.



\end{abstract}
%
%
\section{Introduction}

The mystery of the cosmological constant, or cosmic vacuum, is
probably the most pressing obstacle to significant improving both
fundamental cosmological theories and the models of elementary
particle physics. The problem arises because in the standard
framework of low energy physics, there appears to be no natural
explanation for vanishing or extreme smallness of the vacuum
density. Why is it $10^{-123} M_{Pl}^4$? There is another aspects
of the problem: why is the vacuum density comparable to the
present matter density in the universe? In seeking to resolve
this problem, one naturally wonders if the real world may somehow
be interpreted in terms of symmetry that can relate vacuum to
matter.

This must obviously be symmetry of non-geometric nature, like internal
symmetry which is widely used in particle physics. Such a symmetry
is not related to
space-time and refer to a  correspondence in the properties of
elementary particles and their families. For instance, there is
internal symmetry between neutron and proton, which reveals in
the fact they behave similarly in strong interactions. These two
particles constitute together a family (doublet) of particles,
and the family is characterized by a quantum number called
isospin, or isotopic spin, which is equal to 1/2, in this case.
Isospin is also attributed to other hadrons, and total isospin is
conserved when particles undergo strong interaction. Internal
symmetries are obviously useful in classifying particles and in
leading to some selection rules in particle transformations.

In cosmology, a special correspondence can be recognized, as I will
show below, in genuine physical properties of the components
of cosmic medium. These components, which are often called now the
{\em forms of cosmic energy}, are vacuum (V), cold dark (D) matter,
baryons (B) and radiation (R), or ultra-relativistic energy,
which includes the
cosmic microwave background (CMB) photons and other possible
light or massless particles (neutrinos, gravitons, etc.) of
cosmological origin. This `cosmic family' of energy forms can be
characterized by a time-independent number, called the
{\em Friedmann number}
hereafter, which is the same, numerically, for each of
the energy forms.

This new phenomenon may indeed be identified as a kind of
symmetry, if one uses a universal operational definition of
symmetry by Weyl (1951). According to Weyl (1951), symmetry
exists, if there is an object which remains unchanged after being
affected by different operations. The numerical value of the
Friedmann  number may be considered as an object of this kind,
since it remains unchanged being calculated for each of the
energy forms in a unified manner. The evaluations of the
characteristic number for different energy forms (at any moment
of time) constitute a set of operations that can be performed
without changing the object.

This {\em cosmic internal symmetry} is associated with a conservation
of the Friedmann number, which remains constant during all the time
when the cosmic energy forms exist in nature.

Cosmic internal symmetry is not perfect;  the Friedmann number is
not exactly, but only
approximately, on the order of magnitude, the same, for the four energy
forms. As Okun (1988) mentions, the concept of symmetry is closely
related to the idea of beauty, and the true supreme beauty needs
some slight symmetry breaking to acquire  a mysterious  and
attractive quality of non-finito...

The origin and physical nature of cosmic internal symmetry are argued to
be due to the interplay between electroweak-scale physics and
gravity in the early universe. This enables to found the
characteristic number in terms of the Planck energy scale
and the electroweak energy scale.

Since cosmic internal symmetry relates vacuum to non-vacuum forms of cosmic
energy, it can really provide a framework which suggests a consistent
solution to both aspects of cosmic vacuum problen mentioned above.

Historically, internal symmetry between baryons and radiation was first
recognized soon after the discovery of the CMB (Chernin 1968).

\section{Friedmann integrals}

There are four major energy forms in
modern cosmological models, and cosmic vacuum is one of them. As
is well-known, the initial idea of cosmic vacuum -- in the form of
cosmological constant -- was suggested by Einstein in 1917, and
after the discovery of the cosmological expansion, he discarded
this idea. But actually Einstein believed
that the ultimate fate of the cosmological constant could have to be
determined by
physical experiments and astronomical observations which would
directly examine the effects of cosmic vacuum in the
real universe. We know now that observational data have recently
provided strong evidence in favor of cosmological constant. The
crucial argument has come from the observations of high-redshift
supernovae (Riess et al. 1998, Perlmutter et al. 1999), and this
has  been also supported by considerations concerning the cosmic
age, the large-scale structure, the CMB anisotropy in combination
with cluster dynamics, the dynamics of the Hubble local flow, etc.
(see Cohn
1998, Carol 2000, Bahcall et al. 2000, Sahni \& Starobinsky 2000,
Chernin 2001 and references therein).

The density of cosmic vacuum has proven to be larger than the
total density of all non-vacuum forms of cosmic energy, which are
dark matter, baryons, and radiation.
The best  fit concordant with the bulk of all observational
evidence today (see the references above) is provided by the
following figures for the densities of the forms of energy
measured in the units of the critical density:
\begin{equation}
\Omega_V = 0.7 \pm 0.1, \;\; \Omega_D = 0.3 \pm 0.1, \;\;
\Omega_B = 0.02 \pm 0.01, \;\; \Omega_R = 0.6 \alpha \times
10^{-4},
\end{equation}
\noindent where $1 <\alpha < 10-30$ is a dimensionless constant
factor that accounts for non-CMB contributions to the relativistic energy.
With the Hubble constant $h_{100} = 0.65 \pm 0.10$, the figures
lead to either an open cosmological model or a spatially flat model.

Note that the accuracy in these figures seems to be somewhat
overestimated here;
perhaps a more reasonable estimate for the observation error may
be  $\pm 0.2$ (if not $\pm 0.3$), rather than the formal figure $\pm 0.1$.
Anyway this is not essential for my discussion now: the cosmic phenomenon
I study is measured at the level of orders of magnitude, not tens percent,
so my considerations are not affected by these details.

Let us turn now to the  Friedmann general solution written for the four
cosmic energies of Eq.1:
\begin{equation}
\int{da (A_V^{-2} a^2 + 2 A_D a^{-1} + 2A_B a^{-1} +
A_R^2 a^{-2} - K)^{-1/2}} = t,
\end{equation}
\noindent were $a(t)$ is the curvature radius and/or a scale
factor of the model, $K= 1,0,-1$, accordingly to the sign of the
 spatial curvature.

 Quantities $ A_D, A_B, A_R$ are constants
 that come as integrals from the Friedmann
`thermodynamic' equation which is equivalent to energy and
entropy conservation relation of dark matter, baryons, and
radiation, respectively, in a co-moving volume:
\begin{equation}
A = [ (\frac{1 + 3w}{2})^{2} \; \kappa \rho a^{3(1+w)}]^{\frac{1}{1 + 3w}}.
\end{equation}

Here $w = p/\rho$ is the constant pressure-to-density ratio for a
given energy form; $w = 0, 0, 1/3$ for dark matter,
baryons and radiation, respectively; $\kappa = 8\pi G/3 = (8\pi/3)
M_{Pl}^{-2}$; the Planck energy $M_{Pl} = 1,2 \times 10^{19}$ GeV.

It is interesting that the integral for vacuum, $A_V$ in Eq.(1), is
also given by the same general relation of Eq.(3), if one puts
there $w= -1$,
though the interpretation in terms of particles does not
obviously work for vacuum. It indicates that the physical sense of
all the four constants cannot be simply reduced to the
energy and entropy conservation or number-of-particle conservation,
generally. In any case, the existence of the integrals means that
there is no energy exchange among all the four forms of cosmic
energy.

The integral for pressure-free matter appeared in the Friedmann
cosmology papers and was denoted as $A$ (see Eq.(8) of his 1922
paper). The integrals of Eq.(3) will be below referred to as the
{\em Friedmann integrals}.

The Friedmann integrals $A_V, A_D, A_B, A_R$ given by Eq.(3) are genuine
constant characteristics for the respective forms of energy during all the
time when each of the energies exists. Each of the energies is
represented in Eq.(2) by its corresponding integral independently from
other components. Being constants of integration, the integrals
are completely arbitrary, in the sense that the Friedmann
equations  provide no limitations on them, except for trivial
ones. From the view point of physics, the integrals are
determined by `initial conditions' at the epoch of the origin of
the forms of energy in the early Universe; at that time, each of
the energies  acquires its own integral as a natural
quantitative characteristic, which is then kept constant in time.
At each moment of time, the Friedmann integrals can be estimated
numerically for the corresponding values of the energy densities and
the function $a(t)$. Empirically, the integrals are evaluated for the
present-day figures of these values.

\section{Symmetry}

For the present vacuum domination epoch, one has from Eq.2 a
well-known particular solution that describes the expansion
controlled by vacuum only:
\begin{equation}
a(t) = A_V f(t), \;\;\; f(t) = \sinh (t/A_V), \;\; \exp (t/A_V),
\;\; \cosh (t/A_V),
\end{equation}
\noindent
for open, spatially flat and close models, respectively. The present-day
($t = t_0 \simeq 15 \pm 2$ Gyr)  value of $a(t)$
estimated with the use of this solution and the observed Hubble constant
is approximately $a(t_0) \sim A_V$, for all the three models.

Now the numerical evaluation of the four Friedmann integrals can easily be
made with the figures of Eq.(1) and with $a \sim A_V$:
\begin{equation}
 A_V = (\kappa \rho _A)^{-1/2} \sim 10^{42} GeV^{-1} \sim 10^{61} M_{Pl}^{-1},
\end{equation}
\begin{equation}
A_D = \frac{1}{4} \kappa \rho_D a^3 \sim 10^{41} GeV^{-1}
\sim10^{60} M_{Pl}^{-1},
\end{equation}
\begin{equation}
 A_B = \frac{1}{4} \kappa \rho_B a^3
\sim 10^{40} GeV^{-1} \sim 10^{59} M_{Pl}^{-1},
\end{equation}
\begin{equation}
A_R = (\kappa \rho _R)^{1/2} a^2 \sim 10^{40} GeV^{-1}
 \sim 10^{59} M_{Pl}^{-1}.
\end{equation}

The integrals (that have the dimension of the length) are
evaluated for the open model with $a(t_0) \sim A_V$. It is
easy to see that the evaluation for the flat (with the scale
factor normalized as in Eq.(4)) and close
models gives rise to similar results, on the order of magnitude.
In the estimation of $A_R$, a conservative value $\alpha = 10$ is
adopted.

As one sees from the set of the relations above,
the numerical values of the four integrals have proven
to be close to each other within two orders of magnitude, and this result
may be summarized in a compact formula (Chernin 2001b):
\begin{equation}
A_V \sim A_D \sim A_B \sim A_R  \sim 10^{60 \pm 1 } M_{Pl}^{-1} .
\end{equation}

If this is not a numerical accident, the coincidence of the four
Friedmann integrals  presents a new epoch-independent phenomenon
in cosmology.

The four Friedmann integrals  exist in the Universe since the epoch at
which the four major forms of energy came to existence
themselves, e.g. at least after $t \sim 1$ sec, and will exist
until the decay of the protons at $t \ge 10^{31-32}$ yrs or/and the
decay of the particles of dark matter. In the beginning of this
time interval, the vacuum density is about forty orders of
magnitude less than the density of R-energy that dominates at
that epoch; but the constants $A_A$ and $A_R$ are as close at $t
\sim 1$ sec as they are in Eq.(9). In future, the scale factor
will change in  orders of orders (!) of magnitude during the
life-time of the proton, and so the densities of D-, B-, and
R-energies, as well as the ratios of these densities to the vacuum density,
will change enormously. But the four constant numbers of Eq.(9)
will remain the same keeping their near-coincidence for all this
future time.

In terms of the Friedmann integrals, none of the energy forms, vacuum
including, looks preferable. While they are
very different in their observational appearance and time behaviour, all
of them are characterized by the same (approximately) genuine constant
quantity $A \sim A_V \sim A_D \sim A_B \sim A_R$, which will be called the
{\em Friedmann number}.
The equality of the energy forms, in terms of the Friedmann integrals,
indicates the existence of a special kind symmetry that
relates vacuum to non-vacuum forms of cosmic energy.

It is clear that such a symmetry does not concern space-time,
 in contrast to, say, isotropy of the 3D space of the Friedmann model or
4D space-time of the de Sitter model. In this sense, this non-geometric
symmetry is similar to internal symmetries in particle physics, for instance
 to symmetry between neutron and proton (mentioned in Sec.1).

This cosmic internal symmetry can also be described in a more
formal way. Indeed,  the Friedmann number is a mathematical
object that appears the same (approximately) being evaluated for
each of the four energies. In this formulation, new symmetry
satisfies the general operational definition of symmetry by Weyl
(1951); the evaluations of the number for each of the energy
forms constitute a set of operations that can be performed without
changing the object.

This symmetry is not perfect, and its accuracy is within a few
percent, on logarithmic scale, according to Eqs.(5-9).

Cosmic internal symmetry is associated with a conservation law, since
the Friedmann number is an epoch-independent constant.

To conclude this empirical part of my discussion, let me note that,
historically,  internal symmetry between two forms of cosmic energy,
namely baryons and radiation, was first recognized (Chernin 1968)
soon after the discovery of the CMB. The significance of this
(even partial) symmetry is
obvious from the fact that the relation
$A_B \sim A_R  \sim 10^{59} M_{Pl}^{-1}$ enables alone to quantify
such things as the baryon-antibaryon asymmetry
of the Universe, the entropy per baryon,
the light-element production in the nucleosynthesis, the epoch
of hydrogen recombination, the present-day temperature of CMB,
etc.

\section{Vacuum energy scale}

In any fundamental unified theory, the physical nature of
cosmic internal symmetry would have to be
explained, and the value of the Friedmann number would have to be
calculable. This seems to be a rather remote goal. Nevertheless,
one may try to identify a clue physical factor that might determine
both symmetry and numerical value of the number.

Let me start this theory part of the discussion with some remarks
on the physical nature of cosmic vacuum.
If one assume that the vacuum energy
is completely due to zero oscillations of quantum fields (as
it was suggested not once since the 1930-s -- see, for instance,
a review in Dolgov et al. 1988), the vacuum density is given by an
integral over all frequencies of the zero oscillations. As is clear,
this would give infinite value for the density. One may cut off the
range of the frequencies, introducing a maximal frequency
$\omega_V$; then the integral takes a form: $\rho_V \sim \omega_V
^4 $ (as it can be seen, for instance, from dimension
considerations). With the observed vacuum density, one finds
that the cut-off frequency $\omega_V \sim 10^{-31} M_{Pl}$.

Then one may try to put this maximal frequency into the cosmological
context and compare it with the rate of cosmological expansion in
the early Universe $1/t \sim (G \rho)^{1/2}$;  here $\rho \sim
\rho_R \sim T^4$, and $T$ is the temperature of radiation which
dominates at that times. From this, one finds that $\omega_V \sim
1/t$ at the epoch, when $T \sim 10^{-16} M_{Pl} \simeq 1$ TeV.
The latter value is close to the electroweak energy scale $M_{EW}$,
and therefore $\omega_V \sim M_{EW}^2/M_{Pl}$.

Accordingly, the {\em vacuum energy scale} $M_V$ may be
introduced:
\begin{equation}
M_V \sim \omega_V \sim G M_{EW}^{2}/l_{Pl} \sim (M_{EW}/M_{Pl})^2 M_{Pl}
\sim 10^{-31} M_{Pl}.
\end{equation}

The vacuum energy scale $M_V$ is much less than the characteristic energy
scales in particle physics; it means that very-low-energy physics is
responsible for the nature and structure of cosmic vacuum. According to
Eq.(10), this very-low-energy physics is associated with electroweak scale
physics and gravity. As is seen from Eq.(10), the energy scale $M_V$
is the gravitational potential energy (in absolute value)
of two masses $M_{EW}$ each separated by the distance of the Planck length:
$M_V \sim G M_{EW}^2/l_{Pl}$.

With the vacuum energy scale, the vacuum density is
\begin{equation}
\rho_V \sim M_V^4 \sim (M_{EW}/M_{Pl})^8 M_{Pl}^4.
\end{equation}

Note that the energy scale of Eq.(10) was mentioned by Antoniadis et al.
(1998); the relation of Eq.(11) was found also in a
field-theoretic model (Arkani-Hamed et al. 2000) -- unfortunately, under
some  arbitrary and rather strong additional assumptions.

\section{Interplay between gravity and electroweak scale physics}

The considerations above suggest that the electroweak scale physics
and also gravity might determined observed density of cosmic vacuum.
This `electoweak-gravity' physics may also be the major mediator between
vacuum and non-vacuum forms of cosmic energies in the processes that might
develop in the early Universe at the epoch of TeV temperatures.
If so, the same processes might be behind cosmic internal symmetry.

In a search for theory equations that could connect the Friedmann
integrals with each other,
let me use a standard freeze-out model to show how -- in principle --
the interplay between gravity and electroweak scale physics
might develop in early universe. In the  model,
non-relativistic dark matter is considered as thermal relic of early
cosmic evolution (see, for
instance, the books by Zeldovich and Novikov 1983, Dolgov et al. 1988,
Kolb and Turner 1990, and especially a recent work by Arkani-Hamed et al.
2000).
In the version discussed below, the model is incomplete:
baryonic energy is not included in it,
 and so baryogenesis at TeV temperatures has to be
studied separately (but, perhaps, not independently) of the model.
As for vacuum, dark matter and radiation,
they will be represented in the model by the corresponding
Friedmann integrals $A_V, A_D, A_R$.

According to the model, for stable (or long-living) particles
of the mass $m$, the abundance freezes out when the temperature falls to
the mass $m$ and the expansion rate $1/t$ starts to win over the
annihilation rate,
$\sigma n$, where the annihilation cross-section $\sigma \sim m^{-2}$. So
that at that moment the particle density is
\begin{equation}
n \sim 1/(\sigma t) \sim m^2 (G \rho_R).
\end{equation}
\noindent
Using Eq.(3) for $ A_D$ and $A_R$ and putting  $\rho_D \sim m n$,
one finds:
\begin{equation}
A_D \sim a(t) m^3 M_{Pl}^{-2} A_R.
\end{equation}
\noindent
One also has at that moment $\rho_R \sim m^4$, and because of this
\begin{equation}
 A_R \sim  a(t)^2 m^2 M_{Pl}^{-1},
\end{equation}
where $a(t) \sim A_V (1 + z)^{-1}$, and $z$ is the redshift
at the freeze-out epoch;
in this way, the vacuum integral comes to the model.

One may specify the underlying fundamental physics, referring
to a special significance of the electroweak scale physics, as it
was mentioned above, and assume that only two fundamental energy
scales are involved in the process, namely $M_{EW}$ and $M_{Pl}$.
If so, it is natural to identify the mass $m$ with the
electroweak energy scale $M_{EW}$.

The further treatment of the model can be carried out in two different
ways.

1. One may use the relation
for the vacuum density in terms of the two fundamental energy
scales, $M_{EW}$ and $M_{Pl}$, as given by Eq.(11).  With this
density, the vacuum integral is
\begin{equation}
A_V \sim  (M_{Pl}/M_{EW})^4 M_{Pl}^{-1}.
\end{equation}

Arguing along this line, one may expect that
the redshift $z$ at the freeze-out epoch may be given by
a simple combination of the same two mass scales:
\begin{equation}
z \sim M_{Pl}/M_{EW}.
\end{equation}

Now the model is described by a system of four algebraic Eqs.(13-16)
(with $m = M_{EW}$) for the four numbers $A_D, A_R, A_V, z$. The solution
of the system is:
\begin{equation}
A_M \sim A_R \sim A_V \sim  (M_{Pl}/M_{EW})^4 M_{Pl}^{-1}.
\end{equation}

Thus, in this treatment of the model (Chernin 2001a,c),
the coincidence of the three Friedmann integrals appears as a direct result
of the freeze-out process mediated by the electroweak scale physics.
This physics determines also the numerical value of the integrals.

2. Treating the same model in another way, one may not use the
relation for vacuum density given by Eq.(11), but instead take
into account the empirical equality of Eq.(9). Then, turning to
Eqs.(13,14), one can put there $A_V \sim A_D \sim A_R \sim A $, where, as
above, $m$ is identified with $M_{EW}$. Now one has the following
solution to the equations of the model:
\begin{equation}
z  \sim  M_{Pl}/M_{EW},
\end{equation}
\begin{equation}
A \sim (M_{Pl}/M_{EW})^4 M_{Pl}^{-1},
\end{equation}
\noindent

Thus, the model in its second version gives the value of the Friedmann
integral $A$ and the corresponding redshift in terms of two fundamental
energy scales.  In this case, the expression of Eq.(11)
for the vacuum density follows directly as a result of the model.

Following Arkani-Hamed et al. (2000) and another recent work by
Kawasaki et al. (2000),  one may introduce  the `gravitational
scale' $M_G$, or the `reduced' Planck scale $m_{Pl}$, instead of the
standard Planck scale: $M_G \simeq m_{Pl} \simeq g M_{Pl}$, where
$g \simeq 0.1-0.3$. The dimensionless factor $g$ accounts for the
fact that the gravity constant $G$  enters the exact
relations in combinations like $ 8 \pi G,  6\pi G$, or $32 \pi
G/3$. Similarly, a few dimensionless factors, like the effective
number of degrees of freedom, etc., may also be included in the
model  -- see again the books mentioned above. With this minor
modification, one gets finally:
\begin{equation}
 A \sim g^4 (M_{Pl}/M_{EW})^4 M_{Pl}^{-1} \sim 10^{61 \pm 1} M_{Pl}^{-1},
\;\;\; w = [-1, 0, 1/3].
\end{equation}
\noindent
A quantitative agreement with the empirical result of Eq.(9) looks
satisfactory here.

Similarly, the vacuum density is
$\rho_A \sim g^8 (M_{Pl}/M_{EW})^8 M_{Pl}^4 \sim 10^{-122 \pm 2} M_{Pl}^4$.

The numerical value of the redshift
$z \sim g M_{Pl}/M_{EW} \sim 10^{15}$, and so the
temperature at the freeze-out is $T \sim 1$ TeV $\sim M_{EW}$, which
reflects once again the central role of the electroweak energy scale
in the physical processes involved in the interplay between vacuum and
the non-vacuum forms of cosmic energy.

To summarize the results of the model in its both versions, the freeze
out shows how -- at least, in
principle -- the nature of the symmetry of cosmic energy forms
can be treated in terms of fundamental physics.
It demonstrate also that the Friedmann number (and therefore the vacuum
density) might indeed be calculable in electroweak-gravity physics.

\section{Cosmic coincidence problem}

There are two problems that have been reasonably recognized as
the most severe challenge to the current concepts in cosmology and
theoretical physics. These are the cosmic coincidence problem and
the naturalness problem. Let me show that cosmic internal
symmetry provides a framework, within which a consistent solution
to both problems can be suggested. The two problems can be
treated as two inter-related `secondary' aspects of a more
fundamental phenomenon of symmetry.

The cosmic coincidence problem is seen from Eq.(1): the density of cosmic
vacuum is near-coincident
with the density of dark matter and also with the densities of the
two other forms of non-vacuum energy. Why this is so?

Indeed, the four energy forms are very different in their physical
structure, appearance and time behaviour; vacuum produces acceleration
of the cosmic expansion, while the three other forms produce
deceleration; dark matter is non-luminous, while baryons are visible;
dark matter and baryons are non-relativistic, while radiation is
ultra-relativistic, etc.

One possible approach to the problem would be to assume that
the acceleration of the expansion is produced by non-vacuum energy  which
has negative pressure and negative effective gravitating
density (Peebles and Ratra 1988, Frieman and Waga 1998, Caldwwell and
Steinhardt 1998, Caldwell et al. 1998, Zlatev et al. 1999, Wang et al. 2000).
This energy form, called quintessence,
can naturally be realized in some scalar field models in which the field
depends on time, and so the density of quintessence is diluted with
the  expansion. If so,  the densities involved  in the observed cosmic
coincidence are all diluted, which makes them seemingly more similar
to each other. Quintessence density might temporarily or even always be
comparable
to the density of dark matter. Although such an idea may make the
closeness of all cosmic densities natural, it does not explain the
coincidence that the quintessence field becomes settled with a finite energy
density comparable to the matter energy density just now (see, for instance,
Arkani-Hamed et al. 2000, Weinberg 2000).

The analysis of Secs. 2,3  has led to the recognition of the
tetramerous coincidence of time-independent constant numbers --
in contrast to the epoch-dependent (and therefore accidental, in
this sense) coincidence of the densities. It is clear from the
analysis that the cosmic densities are coincident at the present
epoch simply because this is the epoch of the transition from the
decelerated matter dominated expansion to the accelerated vacuum
domination expansion. Because of cosmic internal (and `eternal')
symmetry, the densities must be coincident just now, because
$a(t) \sim A_V$ at present:
\begin{equation}
\rho_V \sim \frac{M_{Pl}^2}{A_V^2}, \; \;\;\; \rho_D \sim
\frac{A_V^3}{a^3} \frac{M_{Pl}^2}{A_V^2}, \; \;\;\; \rho_B \sim
\frac{A_V^3}{a^3} \frac{M_{Pl}^2}{A_V^2}, \; \;\;\; \rho_R \sim
\frac{A_V^4}{a^4} \frac{M_{Pl}^2}{A_V^2}.
\end{equation}

Another question is why do we happen
to live at a transition epoch? This is among the matters that may
be discussed with antropic principle (see, for instance, Weinberg
1987, 2000).

Are there any other forms of cosmic energy with the equation of
state $p = w \rho$ in the same family of energies characterized by
the known A-number?

If yes, their current densities must be
\begin{equation}
 \rho \sim (1+3w)^{-2} (M_{Pl}/A)^{2} \sim (1+3w)^{-2} \rho_V.
\end{equation}

These densities may be both comparable to or very different from
the four observed densities.
For instance, quintessence, a completely hypothetical form of energy
with equation of state $p = w \rho$,
where $-1 < w < -1/3$, would be described by the fifth(!) energy term in
the Friedmann formula of Eq.(2) and have a density of the order of
the observed vacuum density, if $w$ is not too close to -1/3.
If $w$ is very close to -1/3,
the corresponding density will be much larger, on the order
of magnitude, than the observed densities.

Another extreme case is $w \rightarrow \pm \infty$. In this case,
the corresponding density would be very low.
 As it was demonstrated (Chernin et al. 2001), General Relativity allows,
in principle, `density-free' energy forms with $w = \pm \infty$.
These extreme energy forms
 do not play in cosmology because of the high space-time symmetry
 of the models, but
 they can produce a rather `ordinary' non-uniform space-time with
 spherical symmetry.

Finally, are any other energy families with different Friedmann
numbers possible in cosmology? No way to approach this question
is yet seen...

\section{Naturalness problem}

The observed density of vacuum conflicts drastically with simple
field theoretic expectations. Field theoretic arguments may
explain why the vacuum density should be either zero or
Planckian, but there is no explanation for a non-zero, tiny and
positive vacuum density. This is the well-known `naturalness
problem in theoretical physics' (Weinberg 1989, Okun 1988).

In the framework of cosmic internal symmetry, the
constant characteristic of physical vacuum $A_V$ finds its
symmetrical non-vacuum counterparts in the integrals $A_D, A_B,
A_R$ which are also constants. In this framework, cosmic vacuum looses its
uniqueness and appears among other cosmic energy forms as an
ordinary, `regular' member of the family of cosmic energies. Therefore
the observed value of the vacuum density is seen quite `natural',
because it is this value that yields to cosmic internal symmetry.

On the contrary, the Plank density would look quite `unnatural' in
this context, since the corresponding integral would be much
smaller (in 60 orders of magnitude) than the three other
integrals. Similarly, zero density for vacuum would give infinite
integral...

If cosmic internal symmetry is really a phenomenon of fundamental
level, one may say that the symmetry implies that the observed
density of vacuum $\rho_V \sim \frac{M_{Pl}^2}{A^2}$.

>From the theory point of view, the key point is that the characteristic
 Friedmann number $A$ is calculable within a simple model (Sec.5)
 in terms of the Planck
energy and the electroweak energy scale. As it was demonstrated
above: $A \sim (M_{Pl}/M_{EW})^4 M_{Pl}^{-1}$. With this
 Friedmann number, cosmic internal symmetry sets the vacuum density to its
observed value: $\rho_V \sim 10^{-120} M_{Pl}^4$.

Other attempts to solve the naturalness problems are critically reviewed by
Sahni \& Starobinsky (2000); see also recent works by
Weinberg (2000), Guendelmann (1999), Mongan (2001), Kaganovich (2001),
Tye et al.(2001).

\section{Summary}

>From the earliest days of natural philosophy, symmetry has
furnished insight into the laws of physics and the nature of the
universe. Modern cosmology involves notion of symmetry in a
fundamental way. Geometric symmetries of cosmological models is the
basic element of current knowledge, and the most important recent
discovery in cosmology is related to the symmetry of the 4D
space-time of the observed universe: if cosmic vacuum dominates in
the dynamics of the cosmological expansion, the 4D geometry is
described (approximately) by de Sitter model with its perfect 4D
isotropy.

The same observational
evidence that has led to the conclusion about geometric symmetry
suggests also the existence of non-geometric symmetry in the
universe. New symmetry is related to not space-time, but to its
energy content. This cosmic internal symmetry
reveals a special correspondence in the genuine physical properties of
physical vacuum and non-vacuum forms of cosmic energies, which are
dark matter, baryons and radiation.
The four energies look very different from each other and
behave differently in space and time, but all of them are
characterized by a common genuine physical quantity, called
the Friedmann number. The characteristic Friedmann number is eternal,
practically. It conserved in time during all the epoch, in which vacuum,
dark matter, baryons and radiation are present in nature.

Examining the physical nature of cosmic internal
symmetry leads to the conclusion that symmetry might be due to the
interplay between
gravity and electoweak scale processes developed in the early
universe at TeV temperatures. This way the numerical value
of the Friedmann number proves to be calculated in terms of the
elecroweak energy scale and the Planck energy.

The concept of cosmic internal symmetry demonstrates its
productivity in application to the most challenging problems of
cosmology and fundamental theory. The cosmic coincidence problem and
the naturalness problem are both
understood now in the framework of symmetry which is the more fundamental
physical phenomenon. Cosmic
internal symmetry implies that the observed densities of cosmic energy
forms {\em must} be nearly
coincident at the present epoch, and the vacuum density {\em must} be
at its observed level since the early stages of the cosmological
expansion and practically forever.

Interesting new grounds for further discussion of cosmic internal
symmetry can be provided by recent ideas (see, for instance, Maartens 2000)
of brane-world scenaria; this issue will be discussed in a separate paper.

The work was partly supported by the grant of the Academy of Finland
`Galaxy streams and dark matter structures'.
\vspace{0.5cm}

\section*{References}

Antoniadis, I., Dimopoulos, S., Dvali, G., 1998, Nucl. Phys. B516, 70.

Arkani-Hamed, N., Hall, L.J., Kolda, Ch., H. Murayama, H., 2000.
Phys. Rev. Lett. 85, 4434.

Bahcall, N., Ostriker, J., Perlmutter, S., and Steinhard, P.,
1999. Science 284, 1481.

Caldwell, R.R., and Steinhardt, P.J., 1998. Phys.Rev. D 57, 6057.

Caldwell, R.R., Dav\'e, R., and Steinhardt, P.J., 1998. Phys.Rev. Lett.
80, 1582.

Carol, S., 2000, astro-ph/0004075.

Chernin, A.D., 1968. Nature (London) 220, 250.

Chernin, A.D., 2001a, Physics-Uspekhi, 71, No.11.

Chernin, A.D., 2001b, astro-ph/0101532.

Chernin, A.D., 2001c, astro-ph/0107071.

Chernin, A.D., Santjago D., Silbergleit A.S., 2001, astro-ph/0106144.

Cohn, J., 1998, astro-ph/9807128.

Dolgov, A.D., Zeldovich, Ya.B., Sazhin, M.V., 1988.
{\em Cosmology of the Early Universe.}
(In Russian; Moscow Univ. Press, Moscow (English Ed. {\em Basics of
Modern Cosmology, Editions Frontiers, 1990}.

Frieman, J.A., Waga, I., 1998. Phys. Rev. D 57, 4642.

Guendelman, E.I., 1999, IJMP 14, 3497.

Kaganovich, A.B., 2001, Phys.Rev. D63, 025022.

Kawasaki, M., Yamaguchi, M., Yanagida, T., 2000. Phys. Rev. Lett. 85, 3572.

Kolb, E.W., Turner, M.S., 1990.
{\em The Early Universe.} (Addison-Wesley, Reading).

Maartens, R., 2000, Phys. Rev. D62, 084023.

Mongan, T., 2001, gr-qc/0103012.

Okun L.B., 1988, {\em Physics of Elementary particles}, Nauka, Moscow

Peebles, P.J.E., Ratra, B., 1988. Astrophys. J. Lett. 325, L17.

Perlmutter, S. {\it et al.}, 1999.  Astrophys. J. 517, 565.

Riess, A.G. {\it et al.},  1998. Astron. J. 116, 1009.

Sahni, V., Starobinsky, A., 2000, IJMP 9, 373.

Tye, S.-H.H., Wasserman, I., 2001, Phys. Rev. Lett. 86, 1682.

Wang, L., Caldwell, R.R., Ostriker, J.P., and Steinhardt, P.J., 2000.
Astrophis. J. 530, 17.

Weinberg, S., 1987. Phys. Rev. Lett. 61, 1.

Weinberg, S., 1989. Rev. Mod. Phys. 61, 1.

Weinberg, S., 2000, astro-ph/0005265.

Weyl, H., 1951, {\em Symmetry}, Princeton Univ. Press, Princeton.

Zeldovich, Ya.B., Novikov, I.D., 1983. {\em The Structure and Evolution of the
Universe.} (The Univ. Chicago Press, Chicago and London).

Zlatev, I., Wang, L., Steinhard, P.J., 1999. Phys. Rev. Lett. 82, 896.

\end{document}